\documentclass[nofootinbib, twocolumn, floatfix]{revtex4-1}

\usepackage{graphicx}

\begin{document}
 \title{Quantifying the Fermi paradox in the local Solar neighborhood}

\author{Daniel Cartin}
\email{cartin@naps.edu}\affiliation{Naval Academy Preparatory School, 440 Meyerkord Avenue, Newport, Rhode Island 02841-1519}

\date{\today}
\begin{abstract}

The Fermi paradox highlights the dichotomy between the lack of physical contact with other civilizations and the expectation that technological civilizations are assumed likely to evolve in many locations in the Milky Way galaxy, given the large number of  planetary systems within this galaxy. Work by Landis and others has modeled this question in terms of percolation theory and cellular automata, using this method to parametrize our ignorance about possible other civilizations as a function of the probability of one system to colonize another, and the maximum number of systems reachable from each starting location (i.e. the degree in the network used for percolation). These models used a fixed lattice of sites to represent a stellar region, so the degree of all sites were identical. In this paper, the question is examined again, but instead of using a pre-determined lattice, the actual physical positions of all known star systems within 40 parsecs of the Solar System are used as percolation sites; in addition, the number of sites accessible for further colonization from a given system is determined by a choice of maximum distance such efforts can travel across. The resulting simulations show that extraterrestrial colonization efforts may reach the Solar System, but only for certain values of the maximum travel distance and probability of an occupied system further colonizing other systems. Indeed, large numbers of systems may be colonized with either vessels that lack insufficient travel distance to reach the Solar System or else have a colonization probability where they are statistically unlikely to reach us.

\end{abstract}

\maketitle

\section{Introduction}
\label{introduction}

One of the most engaging questions today is that of the Fermi paradox, resulting from the variance between the observed fact that the Solar System has apparently not been visited by craft from a space-faring civilizations, and the expectation that the large number of potential sites where other life could arise should lead to at least one other alien civilization in the Solar neighborhood. Recent results in the discovery of exoplanets have compounded this, with the realization that planetary systems -- and planets habitable by life similar to the varieties on Earth -- are found around a significant fraction of other stars. As we continue to explore more of the Solar System, the complete lack of extraterrestrial mechanisms within our system becomes more likely. These observations are a continuing source of discussion, and it is likely the true answer to this paradox will not be known until humanity takes definitive steps into the galaxy. At this point, there is only a null result for the existence of other civilizations, from the lack of clear electromagnetic signal transmitted towards the Solar System or proof of star-faring activities. However, by constructing simple models, this lack of evidence can place upper limits on the rate of incidence for these types of civilizations. The need for simplicity in these models is from the large extent of our ignorance about many fundamental questions touched on by the Fermi paradox, such as what conditions are necessary for the advent of (unicellular) life, and for this life to survive a sufficient time to develop complex forms; what is the likelihood of some species of this life to develop a stable technological civilization; and what impetus is required for this civilization to engage in travel to other star systems. This later question is unanswerable at the moment, considering humanity has not taken this final step itself, so at the moment only speculation is warranted. The upshot of this, however, is that when modeling these processes, the model should be as simple as possible, so that its assumptions (both tacit and otherwise) do not influence the conclusions.

In this vein, Landis~\cite{Landis} considered the Fermi paradox by using a percolation model to simulate the process of extraterrestrial colonization. Percolation theory has been used in a wide variety of contexts (see references in \cite{NewZiff}), and offers many accurate descriptions of physical phenomenon with only the barest of inputs. The specific model used by Landis is the following. All possible star systems are simulated as sites on a cubic lattice; one of these is chosen as the initial source of colonization. ``Colonizer" systems will colonize all of their neighbors, with a fixed probability $p$ of these adjacent sites colonizing systems themselves. Those systems that do not become colonizers, with probability $(1 - p)$, will not attempt to colonize any of their neighbors. Whether the system is a colonizer or not is determined at the point of colonization and does not change with time, and systems can only be colonized once. Thus, this model has only two choices of parameters -- the probability $p$ that a system will colonize its neighbors, and the choice of a cubic lattice, i.e. that each system has a total of $N = 6$ neighbors. Because the cubic lattice is a well-studied system in percolation theory~\cite{Stau-Ahar}, it is known there is a critical probability $p_c \approx 0.312$ where the cluster of colonized stars will have essentially infinite range. However, even for probabilities $p \ge p_c$, there exist many ``voids" of uncolonized systems, providing an explanation for the Fermi paradox despite the existence of star-faring civilizations. Wiley~\cite{Wiley} comments on this aspect of Landis' model, pointing out that fairly simple extensions of the model are possible which conceivably negate this result, specifically allowing non-colonizing systems to become colonizers over time, and giving system colonies a finite lifetime, so that they may be re-colonized by their neighbors. These modifications lead to an absence of uncolonized systems or voids in the region of space simulated by the model.

Other work building on that of Landis include Hair and Hedman~\cite{HaiHed}, and Vukoti\'c and \'Cirkovi\'c~\cite{VukCir}. Hair and Hedman consider percolation on a cubic lattice, but allow colonization of nodes diagonal to previously occupied sites; this gives each system in the network a maximum possible degree $N = 26$ (since they examine different choices of accessibility, such as colonization along diagonals in the cubic lattice). In addition, at each time step of their algorithm, every occupied site is allowed to colonize another system with a certain probability. Thus, unlike the Landis model, systems may colonize all, some or none of their neighbors. They assess the physical diameter of the region of occupied systems, showing it scales with the number of algorithm iterations (i.e. physical time elapsed). These authors also note the existence of voids of uncolonized systems, ranging up to thirty light-years in size. On the other hand, Vukoti\'c and \'Cirkovi\'c use a cellular automata model on a two-dimensional square lattice, where the state of each cell represents its level of biological and technological development. Their examination is focused on a broad analysis of astrobiological ideas; however, as a corollary to this, they discuss their findings on the range of clusters representing interstellar civilizations, and thus the implications regarding the Fermi paradox. In particular, they note that the range of sizes obtained for technological civilizations is concentrated at around 100 parsecs (pc), with few examples of larger extent. This implies a range of voids consistent with the results of Landis. Thus, whether the Fermi paradox can be answered by the existence of such voids is an open question.

\section{Percolation model}
\label{percolation}

\subsection{Model assumptions}
\label{assumptions}

\begin{figure}[hbt]
	\includegraphics[width=0.5\textwidth]{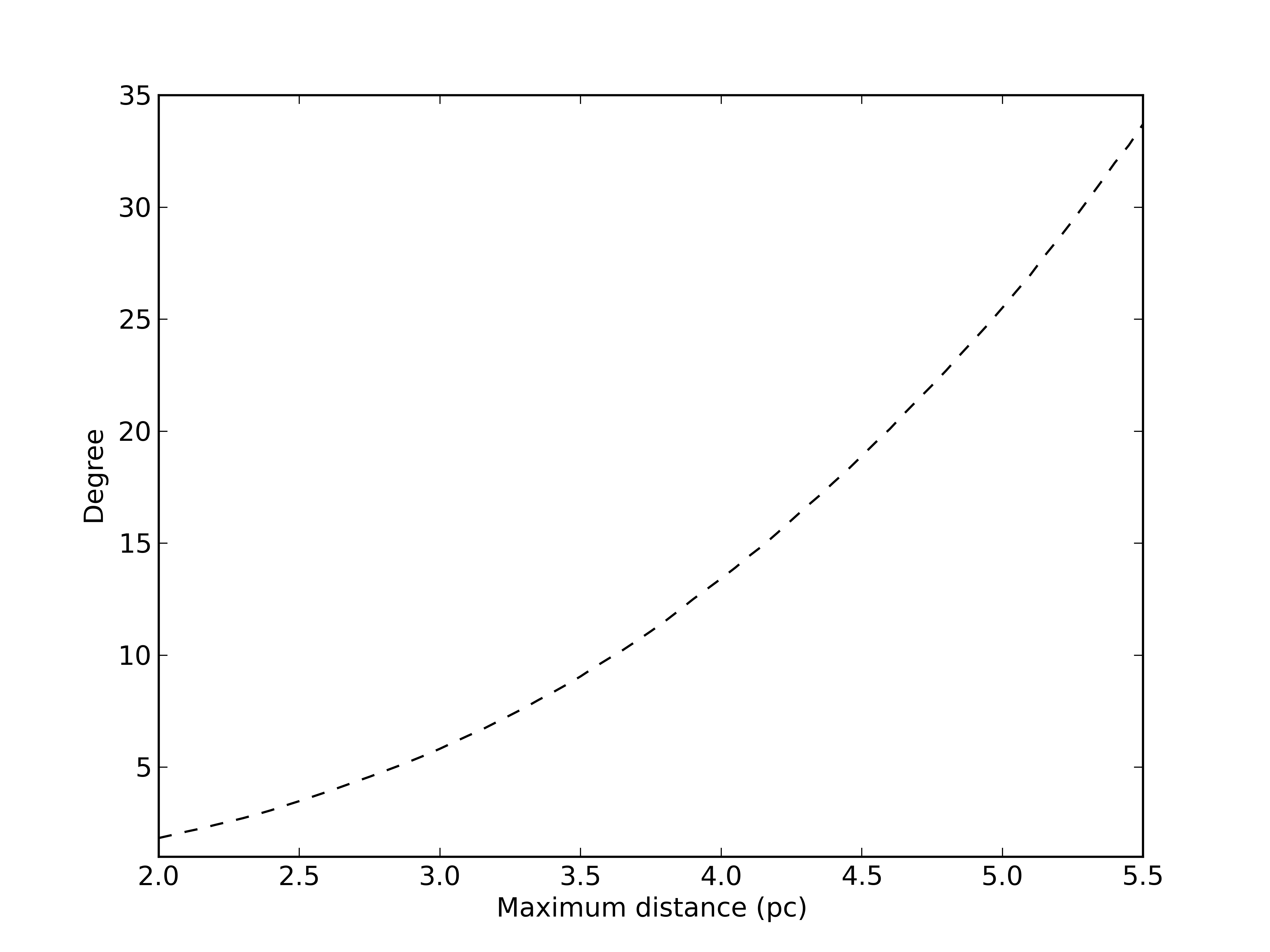}
	\caption{\label{degree}Average degree of all star systems within 20 pc of the Solar System, as a function of maximum distance considered in the graph. Recall that for a cubic lattice (the type used in previous percolation models), the degree for all systems is $N = 6$; this corresponds to a maximum distance of about 3 pc. In the work by Hair and Hedman, the choices $N = 18$ and $N = 26$ were also used, corresponding to maximum distances of 4.4 and 5.0 parsecs, respectively.}
\end{figure}

This paper details a percolation model similar to that of Landis, and explores how the choices of parameters affect the results obtained. As with previous models, one of these parameters is the probability $p$ of successfully colonizing a neighboring system. However the number of neighbors $N$ is not fixed -- instead of simulating these systems as the sites of a cubic lattice, the model uses the actual positions of all stars within forty parsecs of the Solar System as given by the SIMBAD astronomical database~\cite{SIMBAD}. All stars within 0.1 parsec of a neighboring star are grouped together into a single, multiple-star system. Because of this, rather than making an explicit choice of the number of neighbors of each system, e.g. by the use of a specific lattice, the model uses a parameter $D_{max}$ for the largest distance a colonization effort can travel from its initiating star. In other words, the neighbors of a star system are all other systems within a distance $D_{max}$, so that this number of neighbors is not constant for all systems. Thus a choice of $D_{max}$ is equivalent to that of picking the average number of neighbors for each star system. To make this relation explicit, we consider all star systems within twenty parsecs of the Solar System, where it is expected that stellar surveys account for most, if not all, of all existing stars in this vicinity. The variation in the average degree for all sites within this radius as a function of the maximum link distance $D_{max}$ is shown in Figure \ref{degree}. The degree $N =  6$ of a cubic lattice is reached when $D_{max} \approx 3.0$ pc. As shown in previous work~\cite{Cartin}, ships able to travel three parsecs can reach a majority of local star systems, so the choice of a cubic lattice is roughly comparable to what would be expected. There will be some differences, though, since the network of links between star systems less than a distance $D_{max}$ away from each other is a random graph, with a distribution of degrees. This leads to groups of star systems inaccessible to vessels without a certain available travel distance; in other words, the group, as a whole, as zero degree or connections with other systems, even though the average degree of systems overall may be high.

It is worthwhile to consider interpretations of the model parameters at this point. These two parameters $p$ and $D_{max}$ parametrize the sociological and technological barriers, respectively, of a successful colonization effort. For example, $D_{max}$ represents the farthest distance ships may safely travel between systems. This may be dependent on the maximum speed of the ship, how far the ship can travel before critical systems break-down, the lifetime of the species sending the ship (or, with some procedure like suspended animation, how long this lifetime can be prolonged), and other similar factors. Thus, these factors represent either purely engineering features, or else the effectiveness of such solutions to other issues. However, the ability to send a ship is not the same as actually having the motivation to send it; this cultural aspect is quantified in the parameter $p$, the probability of a ship being launched from a given system towards another. As detailed in previous articles (see, e.g. \cite{Millis}), constructing and launching a starship is a substantial effort, even for a civilization using a large portion of the energy and material resources of its home system. In addition, the distances involve mean that there will be little interaction between separate systems, other than by electromagnetic or other light-speed communications.

Based on these considerations, a more statistical view of the parameters $p$ and $D_{max}$ is possible. For example, the maximum distance $D_{max}$ can be based on the probability of colonization craft reaching their destination. As a simple model, suppose the probability of catastrophic failure per distance interval for each craft is a constant value $\lambda$, and this value is independent of the distance already traveled. Since all such craft are likely to have a constant cruise speed during a vast majority of their flight, the probability of failure per time is proportional to that of distance; using a distance scale allows an easier interpretation of the results. The choice of a constant $\lambda$ amounts to an exponential probability distribution for the likelihood of success for each colonization attempt. Using this distribution, the mean expected distance traveled by each craft is $1 / \lambda$, and therefore setting a maximum distance $D_{max}$ is equivalent to choosing a failure probability per parsec for the vessels engaging in the colonization. Similarly, the choice of whether to colonize nearby star systems is likely to vary somewhat over a star-faring civilization, dependent on factors such as the cultural background of the initial population and the abundance of energy and material resources in the source system. Recall that in using all systems within 40 parsecs of the Solar System, there is the inclusion of many brown and white dwarfs, suitable for colonization by a technologically advanced race, but possibly unlikely to have the excess resources to colonize other systems. If the distribution for the outcome of these independent decisions whether or not to colonize again is taken, a probability function of the number of colonizing systems as a percentage of the total number of systems occupied is calculable. This function is likely to be a normal distribution, centered around a mean probability, which is used in the model as the value $p$.

\begin{figure}
	\includegraphics[width=0.5\textwidth]{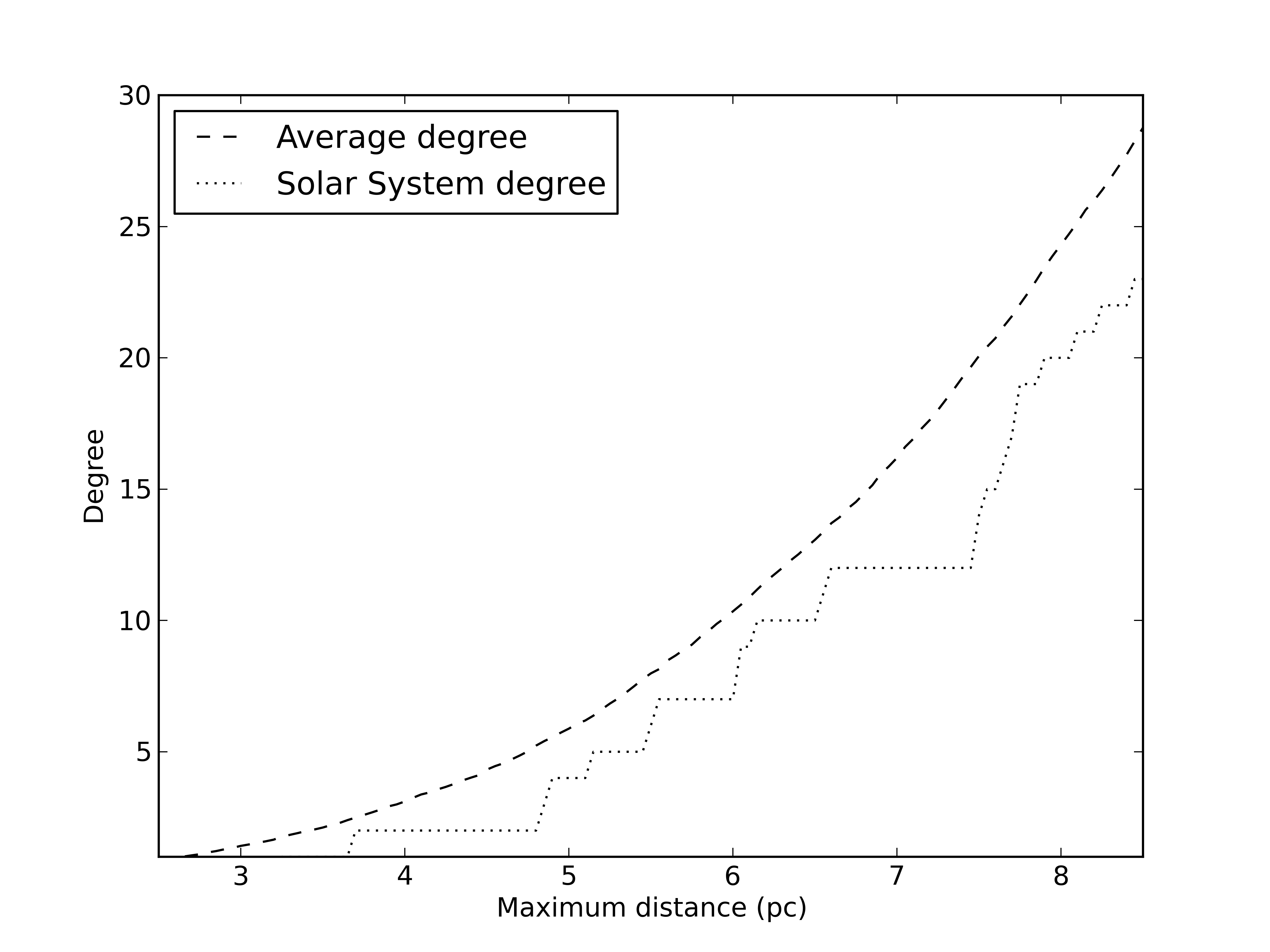}
	\caption{\label{bright-degree}Average degree for systems within 20 pc of the Solar System, as a function of maximum distance $D_{max}$ considered in the percolation network. In this graph, the only systems considered are single stars with spectral types of K or earlier (i.e. ``bright" stars); for reference, the degree of the Solar System is also plotted, which jumps in discrete steps as each new bright star comes within $D_{max}$. For a cubic lattice (the type used in previous percolation models), the degree for all systems is $N = 6$; this corresponds to a maximum distance of about 5 pc, or roughly double the distance required when all systems (not just bright) are considered. In the work by Hair and Hedman, the choices $N = 18$ and $N = 26$ were also used, corresponding to maximum distances of 7.3 and 8.2 parsecs, respectively.}
\end{figure}

One advantage of using actual stellar positions into account is that the distribution of spectral class can be used as well. When Landis made his estimate for $N$, it was only for the number of star systems with certain stellar classes, namely F8 through G9, to be roughly Sun-like, and not in multiple star systems; finding five such stars in the Gliese star catalog within 30 light-years led to an estimate of $N \approx 5$, so that the cubic lattice is a good approximation. A similar division is used here, motivated by astrobiological concerns. Specifically, a primary result of spectral type is whether a planet in the habitable zone is tidally locked to the primary; as seen in Kasting, Whitmire and Reynolds~\cite{KasWhiRey}, the crossover between planets that are locked or not is roughly that between spectral class K and M. Thus, for the purposes of this paper, we divide single stars into ``bright" (spectral class K and earlier) and ``dim" (class M). The class of dim stars includes other types of objects not shining by nuclear fusion, such as solitary white and brown dwarfs, despite the problematic nature of having a habitable zone around these objects~\cite{white-brown}. Finally, since at this point, our knowledge about planet formation in multiple star systems is rather limited~\cite{multiple}, we treat these systems as a separate third category ``multiple". The SIMBAD database provides not only positions for all stars within 40 pc of the Solar System, but for most, the spectral class of each star as well. When a spectral class is not listed, the object is placed in the ``dim" category. This classification leaves the possibilities of using all star systems in the volume under consideration, or else focusing purely on more Sun-like ``bright" stars. If the latter is used, the degree distribution of these stars is different than that for all systems, as reflected in Figure \ref{bright-degree}.

Using only bright stars lends itself to more accuracy. Because dimmer stars tend not to be observed, they are underrepresented in surveys of the local stellar neighborhood, especially for larger distances from the Solar System. This is seen directly in the stellar number density within the 40 pc radius volume used in this work. A reasonable expectation is that this number density $\rho$ would be roughly constant, so that the total number $n_{stars}$ of stars within a volume of radius $r$ would scale as $n_{stars} \sim r^3$. Indeed, this is nearly the case for using only bright stars, where the number density $n_{bright} \sim r^{2.81}$. On the other hand, when all star systems -- i.e. with systems with multiple stars counted only once as a whole -- are taken into account, the scaling of the total number density is $n_{stars} \sim r^{2.13}$. When all stars are counted individually, the number density scales as $n_{stars} \sim r^{2.05}$; the scaling coefficient is less, since there are now more stars at smaller radii. For this reason, the data set containing only ``bright" stars is considered more accurate. From this expectation, Figure \ref{bright-degree} includes the number of neighbors for the Solar System as well for a given $D_{max}$, whereas it is not included when all star systems are used. In the latter data set, the Solar System has a larger degree than average, since more dim stars have been observed close by, while the average degree is lower since there are unobserved stars.

%
%


\subsection{Monte Carlo algorithm}
\label{MC}

The initial motivation for a colonization model based on percolation was to estimate the probability of an extraterrestrial civilization entering the Solar System. Currently, there are no firm ways of stating, e.g. the likelihood of a star-faring civilization emerging at a given star system, so the Fermi paradox must be answered statistically. In other words, all possible clusters of colonized star systems must be simulated, as a function of the colonization probability $p$ and the maximum travel distance $D_{max}$. Here, a ``cluster" refers to a group of systems, all of which are colonized by a single originating site within the cluster, and no other systems are reached from this initial system. For these clusters, several observables are of interest:
\begin{enumerate}

	\item the average size of each cluster
	\item the size of the largest cluster
	\item the average size of clusters containing the Solar System

\end{enumerate}
The specific type of percolation model used is known as {\it site percolation}, where systems or ``sites" are considered either occupied or not (corresponding to whether the matching star system has been colonized). Unlike the results presented in Landis, but the same as that of Hair and Hedman, colonizing systems may settle some but not all of their neighbors. Again, this matches with the statistical interpretation of the colonization probability $p$, where each colonization effort is a choice only of the originating system, not the entire civilization.

In order to quickly calculate the observables previously listed, a Monte Carlo algorithm developed by Newman and Ziff~\cite{NewZiff} is used, and briefly summarized here. A number $M$ of sites is chosen, to represent the star systems of interest within a certain volume, either all of them, or solely the bright stars; for all star systems within 40 pc, $M = 5481$, while for all bright stars, $M = 2182$. A network of links between these sites is created, where each link represents two systems lying within a distance $D_{max}$ of each other. This forms a graph upon which site percolation can occur. For a given observable $Q$ of interest, the algorithm must calculate a value $Q(p, D_{max})$, given any choice of $p$ and $D_{max}$. Considering how to deal with $p$ first, an observable $Q$ as a function of the probability $p$ is given by the binomial expansion
\begin{equation}
\label{binomial}
	Q(p, D_{max}) = \sum_{n=0} ^M  {M \choose n} p^n (1 - p)^{M - n} Q(n, D_{max})
\end{equation}
Here $Q(n, D_{max})$ is the same observable as $Q(p, D_{max})$, but now computed for a network where $n$ sites have been occupied, and the links in the network have a maximum distance $D_{max}$. These observables $Q(n)$ are calculated by a Monte Carlo algorithm, where the network starts with zero sites occupied, and adding one at a time. Thus, instead of considering a cluster initiated at a single site, the algorithm calculates all possible clusters at the same time, without reference to a starting system as the initial colonizer. As $n$ increases, clusters are combined when there is a new link that joins them. At each of these steps, the various observables of interest are found with little computational cost, since $Q(n)$ is frequently a constant-time calculation based on $Q(n - 1)$. The algorithm is run for 1000 trials for each choice of $D_{max}$. From the calculated $Q(n)$, the value of $Q(p)$ can be found for any value of $p$ using the binomial distribution (\ref{binomial}).

\begin{figure}
	\includegraphics[width=0.5\textwidth]{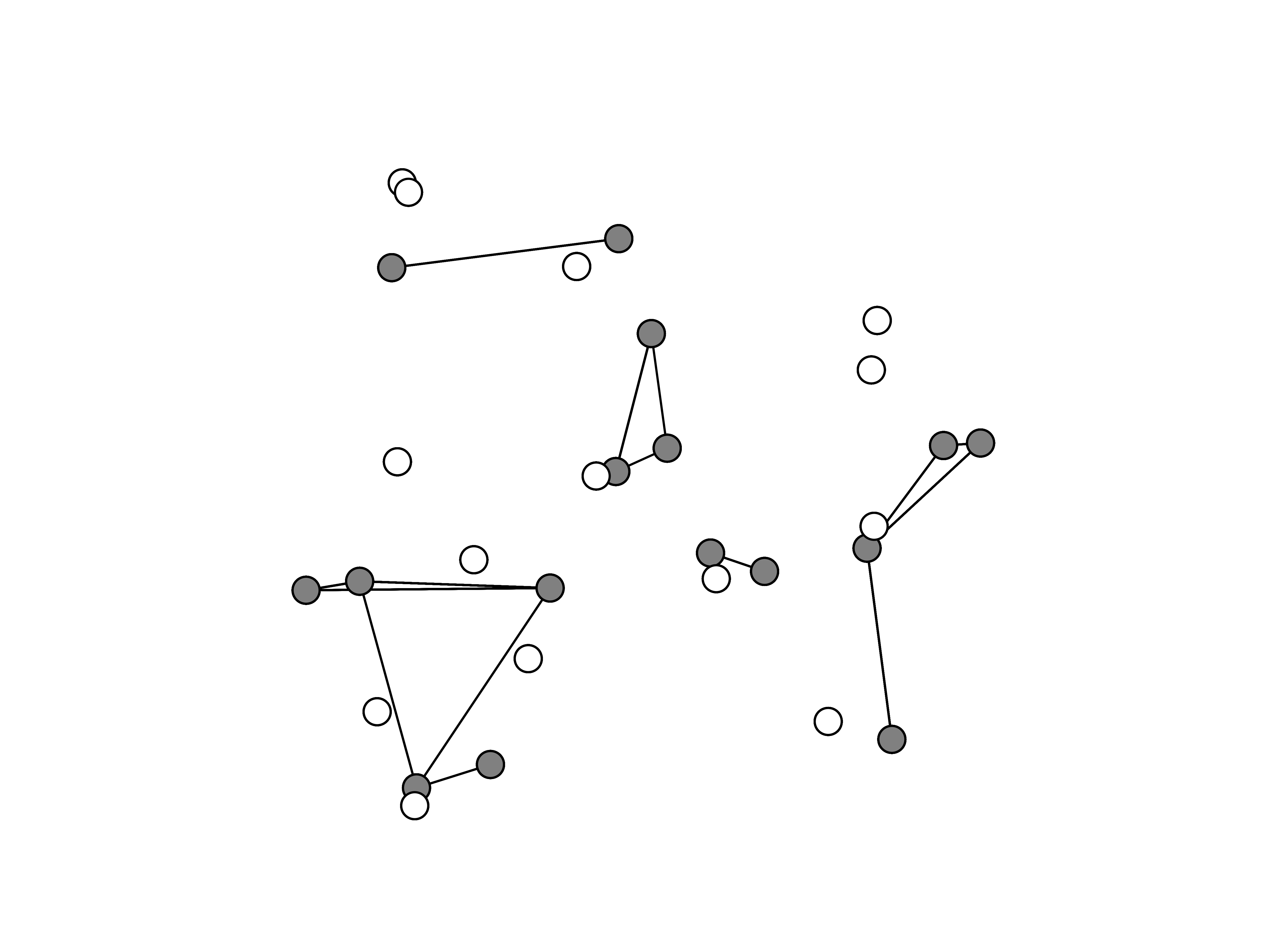}
	\includegraphics[width=0.5\textwidth]{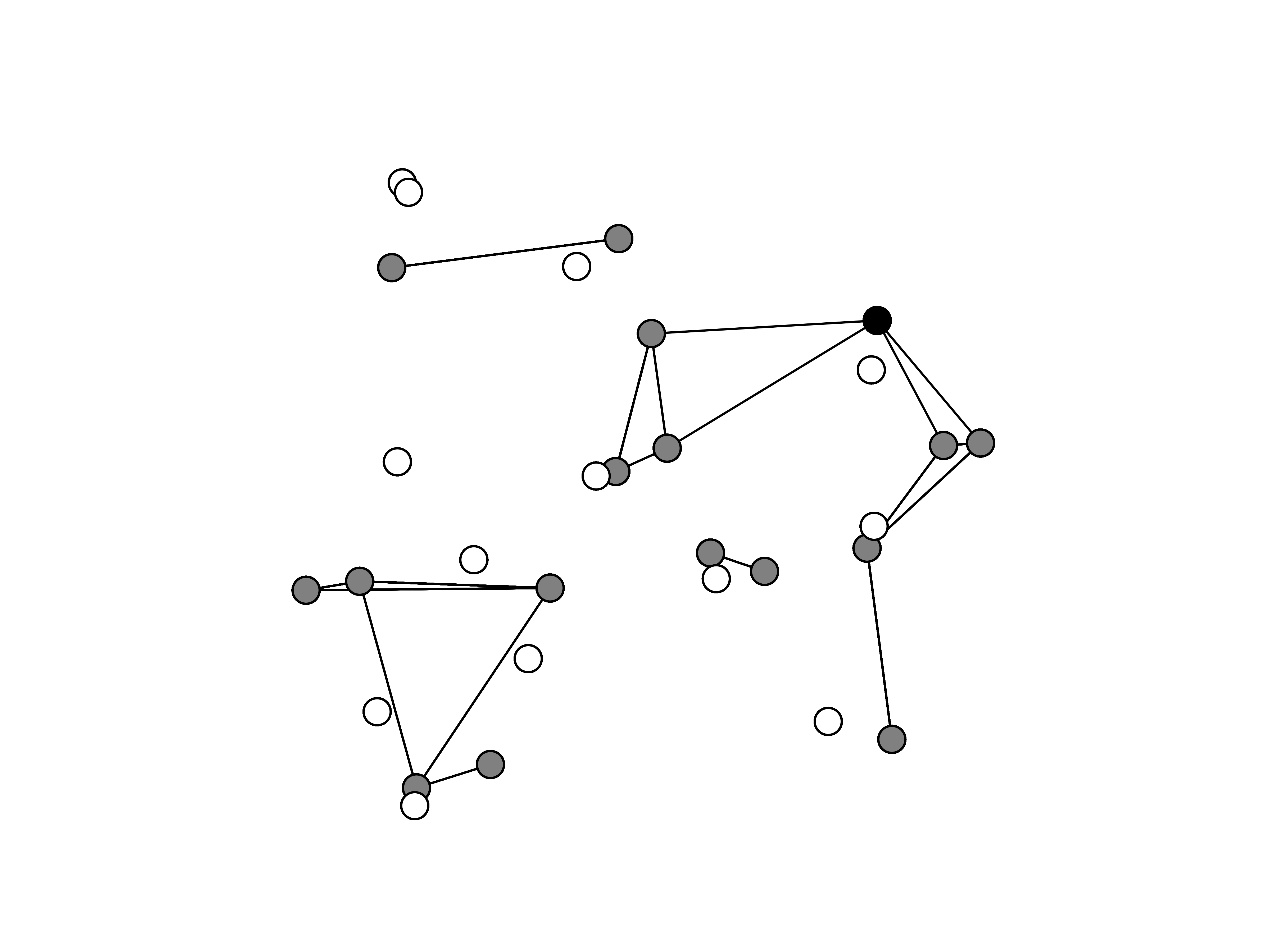}
	\caption{\label{perc-algo}A single step in the Monte Carlo algorithm described in the text. Each node shown represents one of the thirty closest star systems to the Solar System, using their actual positions but projected onto a plane for clarity; occupied (or colonized) nodes are shown in gray, and unoccupied in white. The first picture shows the results of the algorithm after sixteen sites have been occupied, with edges denoting those sites within the same cluster. The second picture shows the process of adding the next node (shown in black), which has the effect of linking two previously disjoint clusters. The remaining unoccupied (white) nodes will be occupied in subsequent steps of the algorithm.}
\end{figure}

An example of a single step in this process is shown in Figure \ref{perc-algo}. These illustrations show the thirty closest star systems to the Solar System as nodes, using their actual positions but projected down into a plane for clarity. The maximum colonization distance $D_{max} = 2.5$ pc. The first shows the result of taking $n = 16$ steps of the algorithm, so that sixteen of the nodes are occupied (or colonized); these are the nodes displayed in gray. The remaining unoccupied nodes are shown in white. Edges are drawn between occupied nodes within a distance $D_{max}$ of each other, so a cluster is represented by sites capable of traveling to each other along the given edges. Thus the nodes of the first graphic are grouped into a total of five clusters. On the next step of the algorithm, another node is randomly chosen to now be occupied; in the second illustration, this is the node given in black. Note that the occupation of this node has now linked together two formerly disjoint clusters, so that after this step (i.e. at $n = 17$), there are now only four clusters. Because of the iterative nature of this algorithm, finding the values of observables $Q(n)$ at each step is straightforward, if they are suitably altered after each addition of an occupied node. For example, if the value of $Q(n = 16)$ is known for the nodes in the first illustration of Figure \ref{perc-algo}, the value $Q(n = 17)$ will only change due to the merger of the two clusters as shown in the second image; contributions of the remaining clusters to the observable $Q(n)$ are unaffected. Thus, each step leads to a short calculation to update the value of $Q(n)$, which then is used in the binomial expansion (\ref{binomial}) to calculate $Q(p)$ for any value of the colonization probability $p$.

In principle, a similar procedure is possible for discreet values of $D_{max}$. Given $M$ total star systems, there are $O(M^2)$ possible links between these systems; these links can be sorted in order of increasing distance. Suppose we start at $D_{max} = 0$, and consider the complete graph of all links with distance $d \le D_{max}$. As $D_{max}$ increases, every time $D_{max}$ passes over the next highest distance in the list of sorted links, each complete graph will add an additional edge. This graph will then remain the same until $D_{max}$ passes over the next link distance in the list. This provides a simple way to compute all possible $Q(p, D_{max})$, in combination with eqn. (\ref{binomial}) above. However, even using only those star systems within 40 pc of the Solar System, this gives over fifteen million different choices of $D_{max}$ to use in the Monte Carlo algorithm, without the gain of much additional insight. Thus, the work described here considers only a small subset of the possible networks determined by $D_{max}$.

\subsection{Results}

\begin{figure}
	\includegraphics[width=0.5\textwidth]{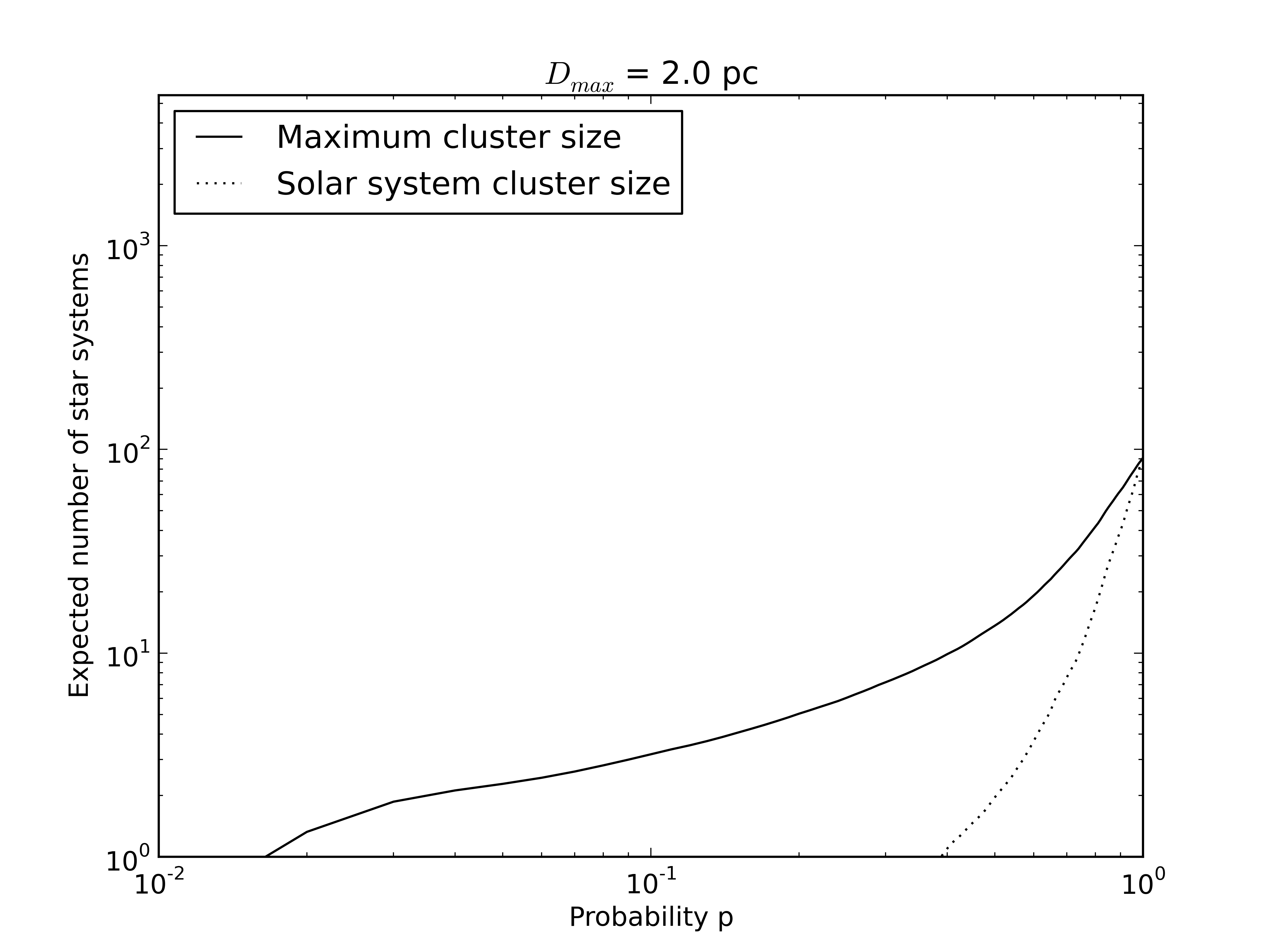}
	\includegraphics[width=0.5\textwidth]{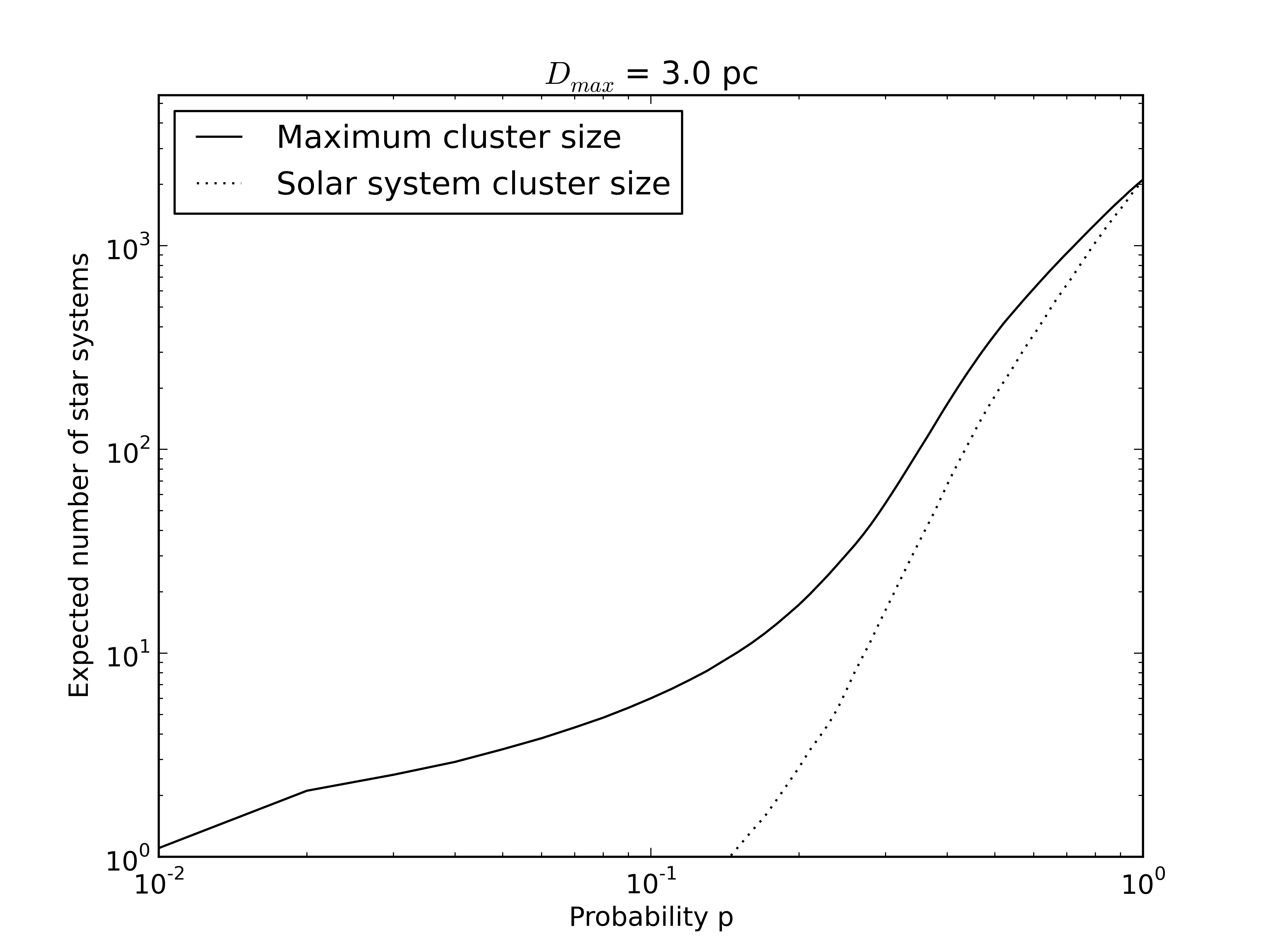}
	\includegraphics[width=0.5\textwidth]{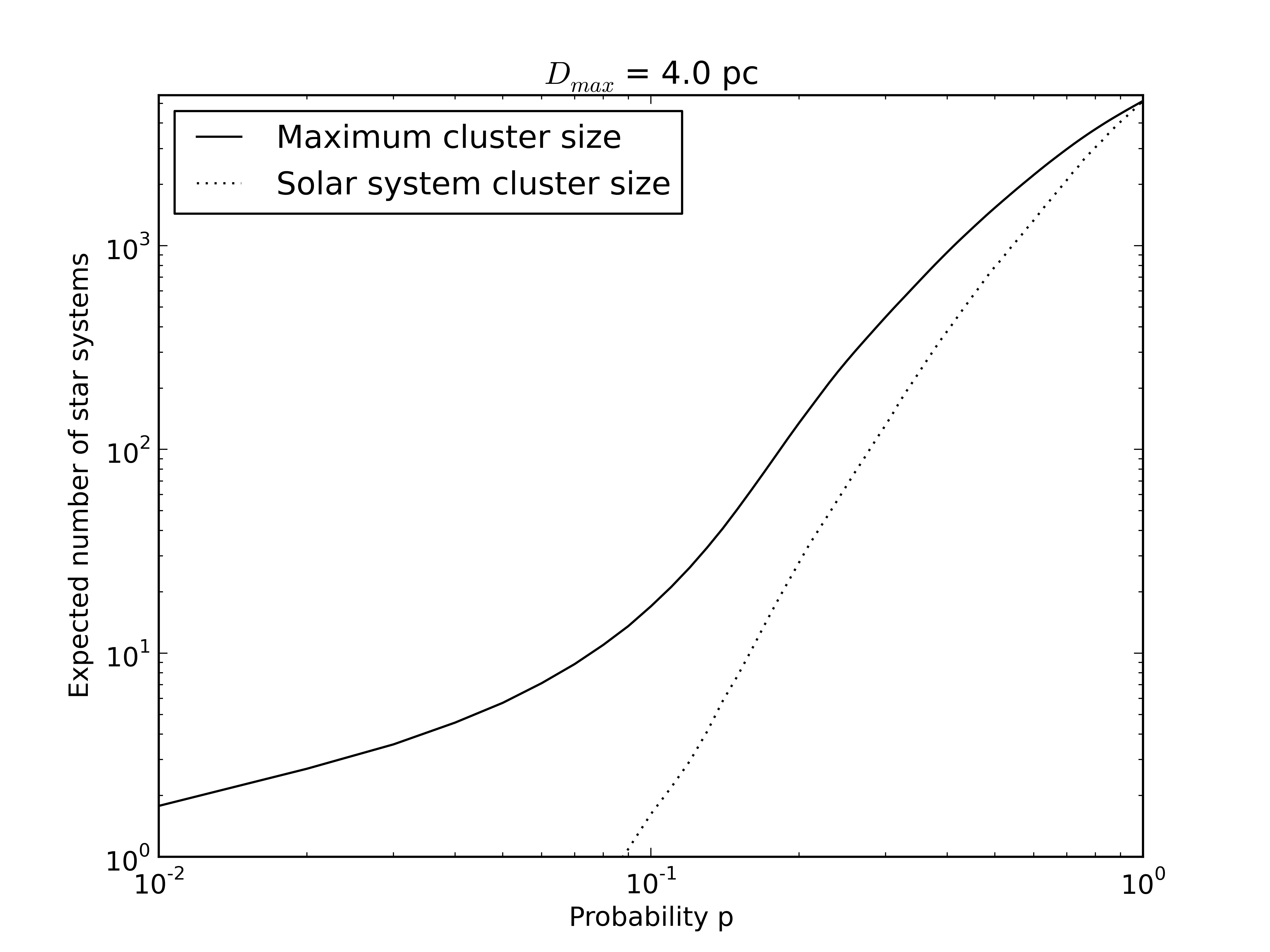}
	\caption{\label{max-vs-Solar-all}Number of systems in the maximum size cluster, and the cluster containing the Solar System, as a function of probability $p$; three choices of maximum travel distance $D_{max}$ are depicted. The maximum distances $D_{max} = 2.0, 3.0$ and 4.0 pc correspond to average degrees $N = 1.84, 5.83$ and 13.4 neighbors for each star system, respectively. For $p = 1$ (i.e. all possible systems are reached), the maximum cluster size is 91 (out of 5481 total) systems for $D_{max} = 2.0$ pc,  2111 systems for $D_{max} = 3.0$ pc, and 5138 systems for $D_{max} = 4.0$ pc. In all three cases presented here, at $p = 1$ the cluster containing the Solar System is the maximum size cluster.}
\end{figure}

\begin{figure}
	\includegraphics[width=0.5\textwidth]{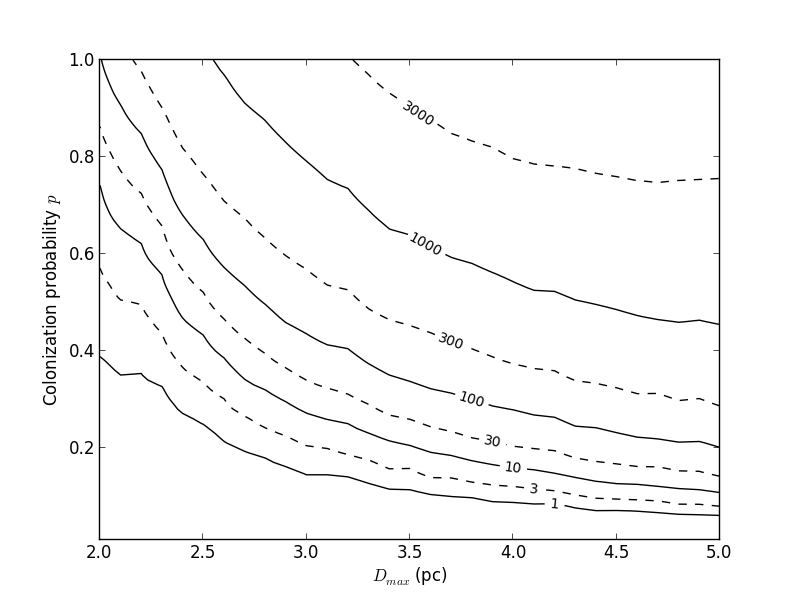}
	\caption{\label{critical-prob-all}Contour plot of the number of systems in the same cluster as the Solar System, as a function of the colonization probability $p$ and the maximum colonization distance $D_{max}$. The contours are selected to be roughly logarithmic in scaling, and the number of systems represented by each contour is shown on the contour line.}
\end{figure}

In Figure \ref{max-vs-Solar-all}, the sizes of the largest cluster and the cluster containing the Solar System have been plotted as a function of the colonization probability $p$ for three choices of the maximum travel distance, namely $D_{max} = 2.0, 3.0$ and 4.0 parsecs. The plots in this figure use all $M = 5481$ star systems within 40 pc. From these graphs, the following interpretations can be made. First, if the size of the Solar System cluster is considered by itself, this gives an idea of the range of star systems that could potentially reach human civilization. Taking the $D_{max} = 2.0$ pc plot from Figure \ref{max-vs-Solar-all}, it is seen that the Solar System is not reached by any star-faring civilization unless $p \ge 0.390$. As $p$ increases beyond this value, the size of the Solar System cluster increased. One interpretation of this is to say that there is a group of star systems -- increasing in size as $p$ increases -- where an extraterrestrial civilization arising in these systems would reach the Solar System after interstellar colonization began. However, continuing with the $D_{max} = 2.0$ pc case with all star systems, even if $p = 1$ (all possible star systems are colonized from the initial site), there are only 91 possible star systems (or 1.66\% of those within 40 pc) from which colonization efforts would reach the Solar System. Obviously this grows as the maximum travel distance $D_{max}$ permitted grows as well. A probability of $p \ge 0.146$ is needed when $D_{max} = 3.0$ pc, leading up to a total of 2111 systems (or 38.5\%) within colonization reach when $p = 1$; the numbers become $p \ge 0.088$ and 5138 systems (or 93.7\%) when $D_{max} = 4.0$ pc. Note that the number of initial colonizer sites never reaches the maximum possible value of $M = 5481$, since there are several systems requiring a larger maximum distance in order to reach a neighboring star system. A contour plot of the number of systems within the Solar System cluster as a function of maximum distance $D_{max}$ and colonization probability $p$ is given in Figure \ref{critical-prob-all}. Increasing $D_{max}$ and $p$ both grow the potential number of systems from which colonization efforts by extraterrestrial civilizations can reach the Solar System.

\begin{figure}
	\includegraphics[width=0.5\textwidth]{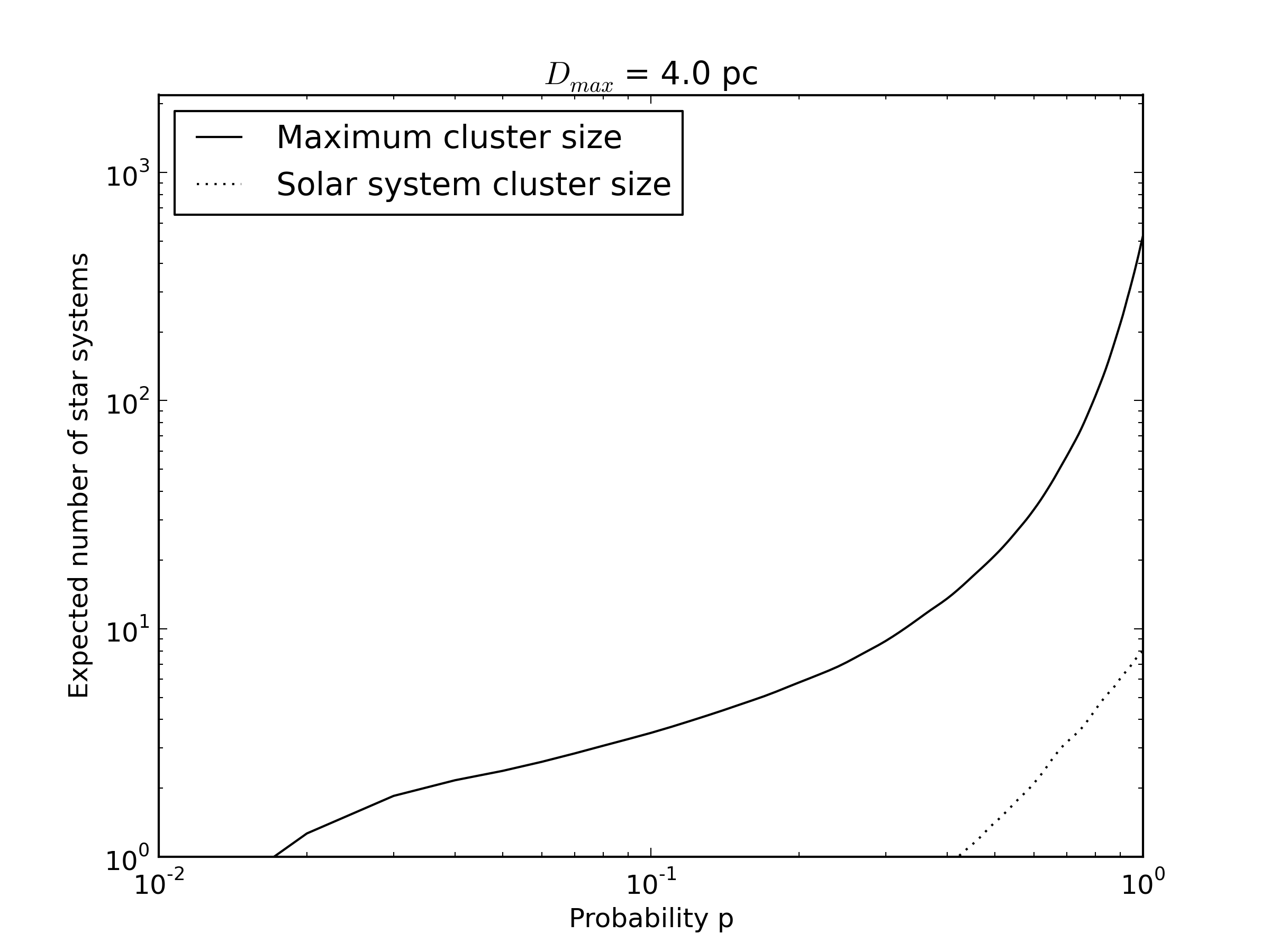}
	\includegraphics[width=0.5\textwidth]{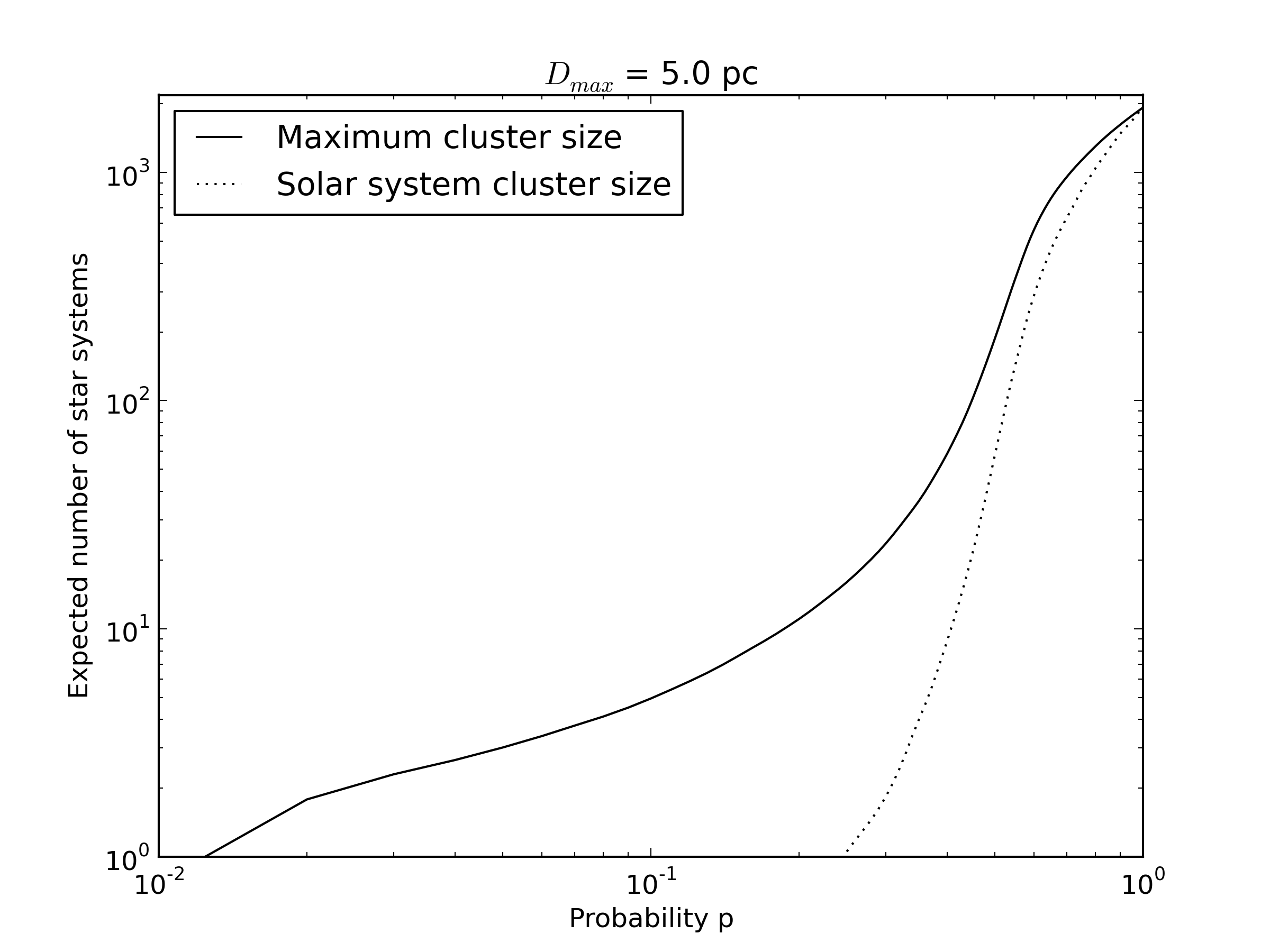}
	\includegraphics[width=0.5\textwidth]{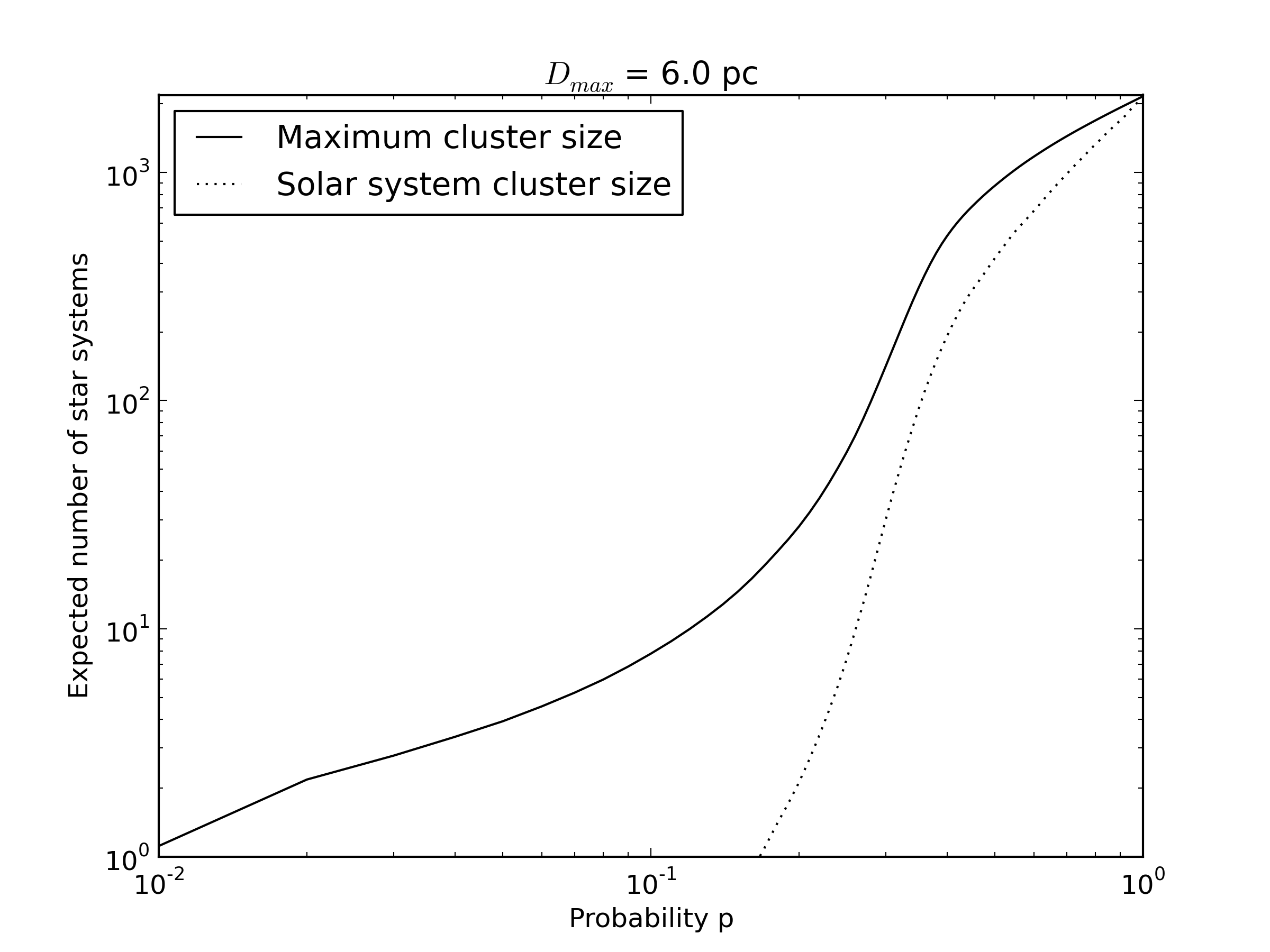}
	\caption{\label{max-vs-Solar-bright}Number of systems in the maximum size cluster, and the cluster containing the Solar System, as a function of probability $p$, including only single stars with spectral class K and earlier (i.e. ``bright" stars); three choices of maximum travel distance $D_{max}$ are depicted. The maximum distances $D_{max} = 4.0, 5.0$ and 6.0 pc correspond to average degrees $N = 3.11, 5.88$ and 10.4 neighbors for each star system, respectively. For $p = 1$ (i.e. all possible bright stars are reached), the maximum cluster size is 528 (out of 2182 possible) stars for $D_{max} = 4.0$ pc, 1926 stars for $D_{max} = 5.0$ pc, and 2158 stars for $D_{max} = 6.0$ pc. The corresponding sizes of the cluster containing the Solar System are 8, 1926, and 2158 stars, respectively.}
\end{figure}

\begin{figure}
	\includegraphics[width=0.5\textwidth]{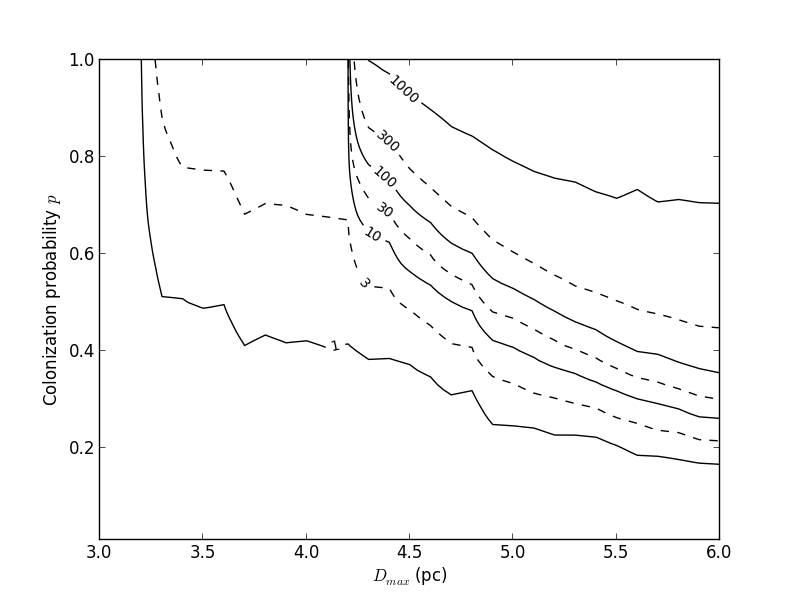}
	\caption{\label{critical-prob-bright}Contour plot of the number of systems in the same cluster as the Solar System, as a function of the colonization probability $p$ and the maximum colonization distance $D_{max}$. Note that $D_{max} \ge 4.3$ pc for this size to be greater than eight stars (including the Sun), and is the same distance cutoff for 50 stars as well. The contours are selected to be roughly logarithmic in scaling, and the number of systems represented by each contour is shown on the contour line.}
\end{figure}

Turning to the case of using only bright stars, seen in Figure \ref{max-vs-Solar-bright}, the necessary probability values to get the same size cluster increase due to the much smaller number of Sun-like stars within the volume under consideration. For example, when $D_{max} = 4.0$, then a probability of $p \ge 0.422$ is required for any colonization efforts to reach the Solar System, and at most only 8 bright stars (or 0.367\% of all bright stars within 40 pc of the Solar System) are within range when $p = 1$; the probabilities and maximum number of possible initial colonization sites becomes $p \ge 0.246$ and 1926 stars (or 88.3\%) for $D_{max} = 5.0$ pc, and $p \ge 0.168$ and 2158 stars (or 98.9\%) for $D_{max} = 6.0$ pc. A contour plot of Solar System cluster size versus maximum distance and colonization probability for only bright stars is given in Figure \ref{critical-prob-bright}. The jaggedness of this contour plot, compared to Figure \ref{critical-prob-all} for all systems, is due to the smaller number $M$ of sites, leading to greater statistical noise and larger jumps in the sizes of clusters as $D_{max}$ increases. Note to have more than eight stars in the Solar System cluster (including the Sun), a maximum distance $D_{max} \ge 4.3$ pc is required. The reason for this is due to the distribution of the spatial positions of the stars used, something not taken into account when a fixed number of neighbors is assumed for each star. Specifically, when $D_{max} = 4.2$ pc, there are a total of seven other bright stars that can be reached from the Solar System. However, for this colonization distance, there are twenty-five other potential clusters that are larger than this, including the maximum size cluster with a total of 922 stars. When the distance threshold is raised to $D_{max} = 4.3$ pc, the Solar System will merge with this largest cluster, which now has 1008 stars within it. This dramatic jump in the number of accessible stars at 4.3 pc is what leads to the striking change in behavior in the contour plot shown in Figure \ref{critical-prob-bright}. Because of the much larger number of sites available when all star systems are included, these large changes in cluster size do not happen in the case of $M = 5481$ shown in Figure \ref{critical-prob-all}.

\begin{figure}
	\includegraphics[width=0.5\textwidth]{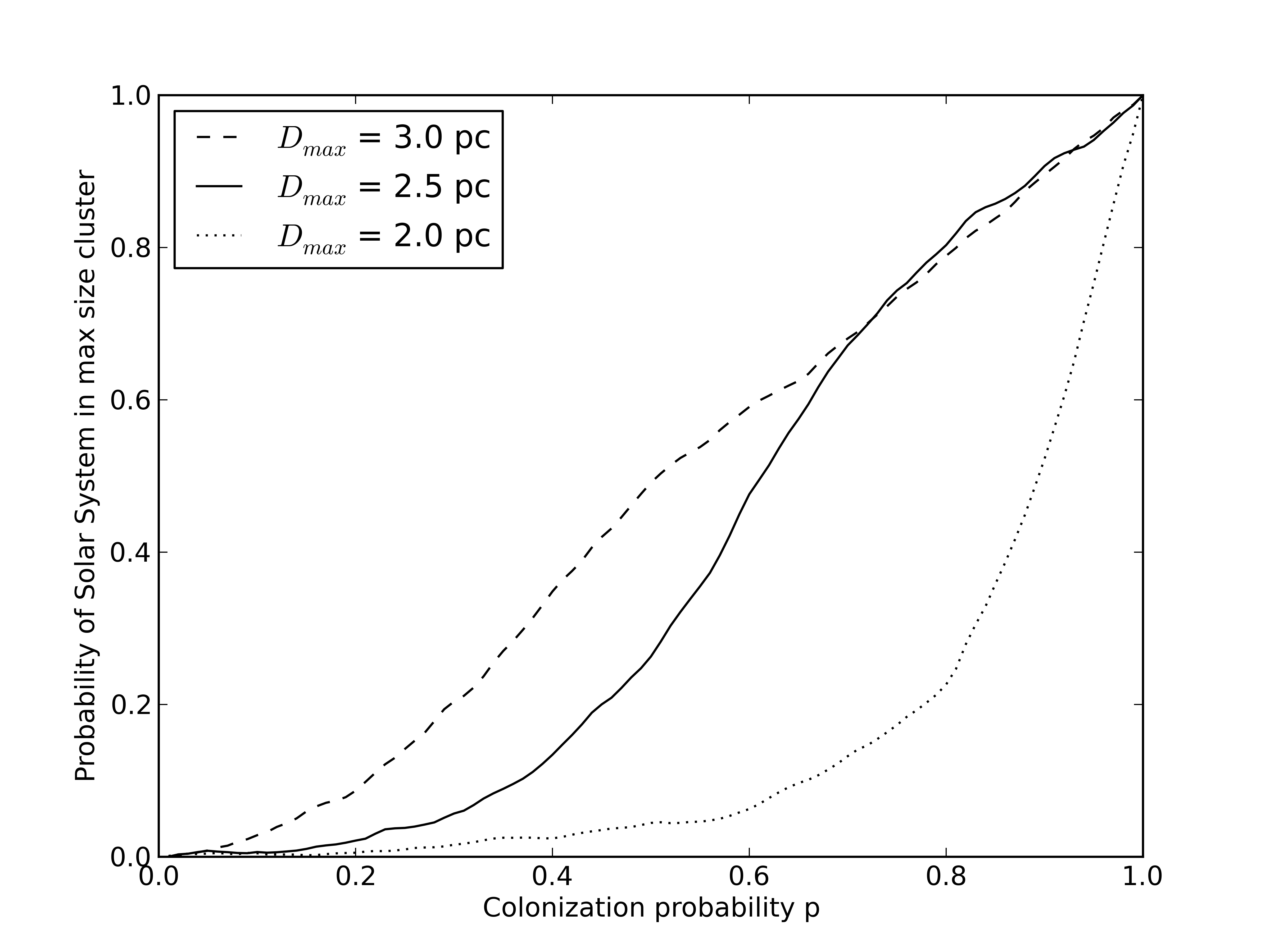}
	\caption{\label{prob-Sol-in-max-all}Probability of finding the Solar System within the maximum size cluster as a function of the colonization probability $p$, for three choices of $D_{max}$ and using all $M = 5481$ systems within 40 pc. Above $D_{max} \approx 3.0$ pc, the probability of the Solar System being in the maximum cluster is approximately linear in the colonization probability $p$.}
\end{figure}

\begin{figure}
	\includegraphics[width=0.5\textwidth]{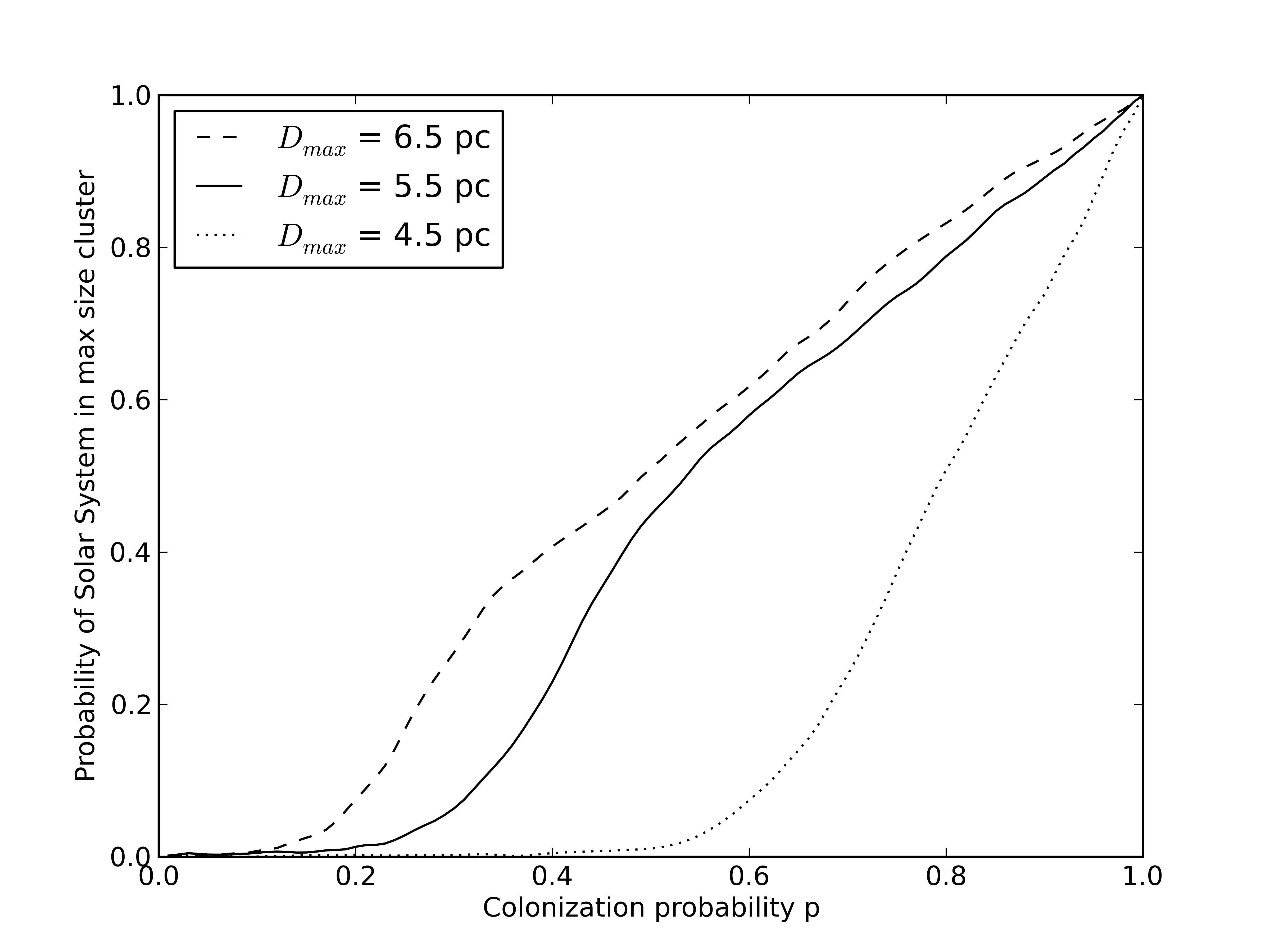}
	\caption{\label{prob-Sol-in-max-bright}Probability of finding the Solar System within the maximum size cluster as a function of the colonization probability $p$, for three choices of $D_{max}$ and using only the $M = 2182$ bright stars within 40 pc. Above $D_{max} \approx 7.0$ pc, the probability of the Solar System being in the maximum cluster is approximately linear in the colonization probability $p$.}
\end{figure}

A second piece of information can be taken from the cluster size versus probability plots in Figures \ref{max-vs-Solar-all} and \ref{max-vs-Solar-bright}, specifically relating the sizes of the maximum cluster versus the cluster containing the Solar System. Note that for all plots, the only time these two sizes are ever the same is when $p = 1$, i.e. all possible star systems are colonized from the original site, and even this is not always true, as seen in the $D_{max} = 4.0$ pc case for only bright stars in Figure \ref{max-vs-Solar-bright}. This gap reflects the chance that an extraterrestrial civilization with a far-ranging colonization reach will never reach the Solar System. Related to this is the fact that significant star-faring programs can begin for rather small colonization probabilities $p$. When $D_{max} = 2.0$ pc and considering all possible target systems, as in Figure \ref{max-vs-Solar-all}, the threshold probability is $p \approx 0.02$ for any kind of colonization cluster; when $p \approx 0.30$ -- the probability needed for any chance of the Solar System being reached -- the maximum cluster has on the order of ten star systems. For larger values of the maximum distance, the required probabilities are below $p = 0.01$. Thus, interstellar civilizations are possible even when the motivation to colonize (as parametrized by $p$) is rather small. Figure \ref{prob-Sol-in-max-all} gives the explicit probability of the Solar System residing in the maximum size cluster as a function of the colonization probability $p$ for three choices of $D_{max}$, when using all star systems within 40 pc of the Solar System; Figure \ref{prob-Sol-in-max-bright} is the corresponding graph when only bright stars are considered.

\subsection{Discussion}
\label{discuss}

Ever since its formal statement, the Fermi paradox has been a question garnering considerable discussion over the years. A wide varieties of possible explanations -- technological, sociological, and otherwise -- have been proposed for its resolution. This paper has used the percolation model of Landis and others to study this issue, using as simplified a model as possible. Thus, the progression of a colonization effort through this volume was parametrized by only two variables -- the probability that an uncolonized star system will be settled by a neighboring colony, and the maximum distance over which this can occur. In other words, the results presented here depend only on broad measures of an extraterrestrial civilization's propensity to migrate to other star systems (their colonization probability $p$), and their technological ability to do so (the maximum distance $D_{max}$). The choice of this maximum distance is equivalent to a choice of the distribution of neighbors for each star system, so using this distance has shown how the number of available neighbors (due to the location of stars) alters the number of the largest colony cluster and the number of systems that could potentially reach the Solar System. Since the spectral classes of all stars are known, percolation could be studied using only bright single stars -- as was chosen by Landis -- as well as taking all star systems in the volume of interest.

The main result of this work is to show that there is credence to the notion that the Solar System is unvisited because it lies in an uncolonized cluster or void. This is most notable when only bright stars are used, but the idea is also valid when all star systems are taken into consideration. For every choice of colonization probability and maximum travel distance, there remained the possibility that an extraterrestrial colonization effort would miss the Solar System. In particular, large-scale clusters of hundreds or thousands of star systems could exist nearby, without reaching the Solar System.  One cause for this is simple physical proximity. An extraterrestrial civilization could colonize almost a thousand bright stars within the spatial volume studied here, for example, by using vessels traveling up to 4.2 parsecs, without having the corresponding ability to reach the Solar System, regardless of their possible wish to do so. In fact, it is conceivable that the size of this cluster could be even larger if the size of the volume considered is extended further. Another cause is the value of the colonization probability. If this is chosen to be a relatively low value, such as $p = 1/3$ as picked by Landis, then solely from mere chance, the Solar System could remain unvisited, regardless of how far colonizing vessels could travel. In fact, rather low values of the colonization probability can lead to significant numbers of colonized systems, especially as the maximum travel distance increases, without the Solar System being part of this cluster. Thus, the apparent lack of alien visits to the Solar System appears to be from two possible factors: (1) other extraterrestrial civilizations have a relatively low desire to colonize other systems, and (2) these civilizations have the technology to visit large numbers of other systems, but these vessels cannot travel sufficient distance to reach our local cluster of stars. Either extraterrestrial civilizations are happy with the number of star systems they have already colonized, or the Solar System may simply lie in a galactic cul-de-sac.

\section{Conclusions}

This paper re-examines the Fermi paradox using percolation theory, along the lines of previous efforts by Landis and others; the range of colonization by an extraterrestrial civilization is modeled by the number of sites occupied in simulated percolation, with a probability parameter giving the likelihood of one system being colonized by another, previously settled system. The new feature considered here is to use the actual physical locations of all star systems within 40 parsecs of the Solar System, rather than assuming that all possible colonization sites are on a uniform lattice. This allows a variation of the maximum possible distance a colonization mission can travel from its originating system, showing the resulting changes in the overall number of colonized systems and, more importantly, the chance that such efforts will reach our Solar System. The methods of Newman and Ziff are used to numerically compute observable quantities for this percolation model for two cases: using all stars within 40 parsecs of the Solar System, and only taking those ``bright'' stars (i.e. spectral class K and earlier). As may be expected, in both scenarios, the number of other systems would have colonization missions reaching the Solar System increases as both the maximum travel distance and colonization probability increase. However, substantial interstellar civilizations can exist even when these two parameters are relatively low (especially when only ``bright'' stars are used), giving rise to voids comprised of uncolonized systems, in agreement with Landis' original resolution of the Fermi paradox.

\end{document}